\documentclass[preprint,prc,aps,showpacs,showkeys,groupedaddress,floatfix]{revtex4-1}
\usepackage{epsfig}
\usepackage{dcolumn}
\usepackage{bm}
\usepackage[latin5]{inputenc}
\usepackage{graphics}
\usepackage{graphicx}
\usepackage{epsfig}
\usepackage{amssymb}
\usepackage{amsmath}
\usepackage{color}

\begin{document}
\title{On the Role of Differentiation Parameter
\\ in a Bound State Solution of the Klein-Gordon Equation}
\author{B.C. L\"{u}tf\"{u}o\u{g}lu}
\affiliation{Department of Physics, Faculty of Science, Akdeniz University, Dumlupinar Bulvari, 07058
Antalya, Turkey,}

\affiliation{Department of Physics, Faculty of Science, University of Hradec Kr\'{a}lov\'{e}, Rokitansk\'{e}ho 62, 500\,03 Hradec Kr\'{a}lov\'{e}, Czechia.}
\date{\today}
\begin{abstract}
Recently, the bound state solutions of a confined Klein-Gordon particle under the mixed scalar-vector generalized symmetric Woods-Saxon potential in one spatial dimension have been investigated. The obtained results reveal that in the spin symmetric limit discrete spectrum exists, while in the pseudo-spin symmetric limit it does not. In this manuscript, new insights and information are given by employing an analogy of the variational principle. The role of the difference of the magnitudes of the vector and scalar potential energies, namely the differentiation parameter, on the energy spectrum is examined. It is observed that the differentiation parameter determines the measure of the energy spectrum density by modifying the confined particle's mass-energy in addition to  narrowing the spectrum interval length.

\end{abstract}
\maketitle 
\section{Introduction}\label{intro}

In nuclear physics, the spin symmetry (SS) and pseudopsin symmetry (PSS) concepts, originally postulated by Smith \emph{et al.} \cite{RefSmithTassie1971}  and Bell \emph{et al.} \cite{RefBellRuegg1975}, were widely used to explain the nuclear structure dynamical phenomena \cite{RefHaxel1949, RefBohr1982, RefDudek1987, RefByrskietal1990, RefNazarewiczetal1990}. The basics of these symmetries were explored comprehensively  and the conclusion was that they depended on the existence of vector, $V_v$, and scalar, $V_s$, potential energies  \cite{RefBahrietal1992, RefBlokhinBahriDraayer1995, RefGinocchio1997}. In 1997, Ginocchio revealed that PSS and SS occur with an attractive scalar and a repulsive vector potential energies that satisfy $V_v + g V_s= \varepsilon^-$, $V_v - g V_s= \varepsilon^+$ conditions, respectively \cite{RefGinocchio1997}.

The Dirac equation (DE) was investigated by using various potential energies in the SS and PSS limits. For instance in the SS limit, Wei \emph{et al.} obtained an approximate analytic bound state solution of the DE  by employing the Manning-Rosen potential \cite{Wei_et_al_2008_1}, the deformed generalized P\"oschl-Teller potential \cite{Wei_et_al_2009_2} energies. Furthermore, in the same limit, they proposed a novel algebraic method to obtain the bound state solution of the DE  with the second P\"oschl-Teller potential energy \cite{Wei_et_al_2010_4}. In their other work, they examined the symmetrical well potential energy solutions in the DE within the exact SS limit \cite{Wei_et_al_2010_5}.  They contributed the field with important papers in the PSS limit too. For instance, they discussed an algebraic approach in the DE for the modified P\"oschl-Teller potential energy \cite{Wei_et_al_2009_3}. Moreover, they applied the Pekeris-type approximation to the pseudo-centrifugal term and investigated the bound state solutions in the DE for Manning-Rosen potential \cite{Wei_et_al_2010_6} and modified Rosen-Morse potential \cite{Wei_et_al_2010_7} energies.

Another relativistic equation, Klein-Gordon equation (KGE), also has been the subject of many scientific investigations in the SS and PSS limits. Ma \emph{et al.} studied the $D$-dimensional KGE with a Coulomb potential in addition to a scalar potential \cite{Ma_et_al_2004}. Dong \emph{et al.}  obtained the exact bound state solution of the KGE with a ring-shaped potential in the SS limit \cite{Dong_et_al_2006}. Hassanabadi \emph{et al.} studied the radial KGE for an Eckart and modified Hylleraas potential energy in $D$-dimension by using supersymmetric quantum mechanics technique \cite{Hassanabadi_et_al_2012_3}. In another paper, Hassanabadi \emph{et al.}  sought solutions of bound and scattering states on the P\"osch-Teller potential energy in the KGE in the SS limit \cite{Hassanabadi_et_al_2013_1}. In the same year, Hassanabadi \emph{et al.} examined the KGE with vector and scalar Woods-Saxon potential (WSP) energy and  presented the scattering case solutions in terms of hypergeometric functions \cite{Hassanabadi_et_al_2013_2}. Only a while ago, Sargolzaeipor \emph{et al.} extended the KGE in the presence of an Aharonov-Bohm magnetic field for the Cornell potential. Then, they introduced superstatistics in deformed formalism and derived the effective Boltzmann factor  with modified Dirac delta distribution \cite{Sargolzaeipor_et_al_2018}.

Recently, we examined the scattering and bound state solutions of the KGE under the generalized symmetric Woods-Saxon potential (GSWSP) energy in the SS and PSS limits \cite{RefLutfuogluJiriJan2018, RefLutfuogluEPJP2018}. We observed that in the scattering case, the solutions existed in the SS and PSS limits \cite{RefLutfuogluJiriJan2018}.  In the bound state case, unlike the scattering case,  bound state solutions existed only in the SS limit \cite{RefLutfuogluEPJP2018}.

GSWSP energy is the generalization of the well-known WSP energy \cite{RefWoodsSaxon1954} by introducing surface interaction terms \cite{BookSatchler}, and  has been  investigated in many research articles \cite{RefKobosetal1982, RefKobosetal1984, RefKouraYamada2000, RefBoztosun2002, RefBoztosunetal2005, RefDapoetal2012, RefCandemirBayrak2014, RefBayrakAciksoz2015,  RefBayrakSahin2015, RefLutfuogluetal2016, RefCapakGonul2016, RefLiendoCastro2016, RefLutfuogluetalb2016, RefOnyeajuetal2017, RefLutfuogluetal2017, RefLutfuoglu2018, RefLutfuogluCJP2018, RefBCL_Ikot_et_al_2018, BCL_et_al_2018}. In one spatial dimension, it is given in the form of
\begin{eqnarray}\label{gws}
  V_{GSWSP}(x)&=&\theta{(-x)}\Bigg[-\frac{V_0}{1+e^{-\alpha(x+L)}}+ \frac{We^{-\alpha(x+L)}}{\big(1+e^{-\alpha(x+L)}\big)^2}\Bigg]\nonumber  \\
  &&+ \theta{(x)}\Bigg[-\frac{V_0}{1+e^{\alpha(x-L)}}+  \frac{We^{\alpha(x-L)}}{\big(1+e^{\alpha(x-L)}\big)^2}\Bigg]. \label{GSWSP}
 \end{eqnarray}
Here, $\theta{(\pm x)}$  denotes the Heaviside step function. GSWSP energy possesses four parameters. Among them, $V_0$ controls the potential well depth, $L$ adjusts  the effective well length and $\alpha$ determines the well's slope. These three parameters have positive values in this paper to allow an investigation of the bound state solutions in a potential well.  The fourth parameter, $W$ is the additional one to the parameters of the WSP and it determines the surface effects type and magnitude.  Since the GSWSP energy was mathematically derived with the first spatial derivative of the WSP, the $W$ parameter is linearly proportional to the other three parameters. The proportionality constant can be either negative or positive, and it is used to determine whether the surface effect is attractive $W<0$, or repulsive $W>0$ \cite{BookSatchler}.

On the other hand, in classical mechanics, an instantaneous configuration of a system is described by  $N$ generalized coordinates. A set of the generalized coordinates that is composed of both the coordinates and their respective momenta is represented with a point in the phase space. The development of the system over time is expressed by the motion of the point in the configuration space. Consequently, the motion of the system between any two distinct times is defined by the trajectory ($N$-dimensional curve) in the configuration space. Note that, the resultant curve in the configuration space is  called "the configuration space trajectory" and it should not be interpreted as the real trajectory of any particle.

In general, in a configuration space there is an infinite number of configuration space trajectories between two points. All physical systems determine their configuration space trajectory by the Hamilton's principle. Hamilton's principle states that the true evaluation of a system between the initial and final points in the configuration space is described by that configuration space trajectory (among infinite number of all possible configuration space trajectories), along which the action functional has a stationary point.

The Hamilton's principle, sometimes called the variational principle,  has a wide usage in physics, not only in classical mechanics but also, in the theory of relativity, statistical mechanics, quantum field theory, etc. \cite{LandauBook, RefGoldsteinBook}.

In this paper, our main motivation is to investigate the role of the $\varepsilon^+$ parameter, hereby we call it the differentiation parameter,  on the energy spectrum via an analogy of the variational principle. We assume the differentiation parameter is the only generalized coordinate that determines the bound state energy spectrum. We assign different values to the differentiation parameter and we calculate their corresponding spectra. In the analogy of the Hamilton's principle, we assume those spectra are different configuration space trajectories. Then, we investigate the role of the differential parameter on the spectra.

Note that, it is a very well known fact  that  the differentiation parameter is equal to zero in finite range potential wells \cite{RefGinocchio1999}. This value corresponds to configurational space trajectory in the language of the variational approach.

We construct the paper as follows. In section \ref{sec:Model}, we started with the KGE in the SS limit and closely followed the paper \cite{RefLutfuogluJiriJan2018}. We obtained the most general bound state solution in the presence of the differentiation parameter. We discussed first the bound state conditions, and then the continuity conditions. Then, we derived the quantization scheme and we obtained the wave function solution. In section \ref{sec:uygulama}, we employed the Newton Raphson (NR) method to obtain numerical results out of the derived transcendental equations. We discussed the role of the differentiation parameter on the energy spectrum of GWSP wells with repulsive and attractive surface interactions, respectively. In section \ref{sec:sonuc}, we gave a brief conclusion to finalize the paper.

\newpage
\section{The Klein-Gordon equation and Bound state solution}\label{sec:Model}
We start with the KGE in the SS limit \cite{RefLutfuogluEPJP2018}
\begin{eqnarray}
   \Bigg[\frac{d^2 }{d x^2}+\frac{1}{\hbar^2c^2} \bigg[E^2-\big(\varepsilon^+ -m_0c^2\big)^2 -2V\Big(E-\big(\varepsilon^+ -m_0c^2\big)\Big)\bigg] \Bigg]\phi(x) &=& 0.  \,\,\,\,\,\,\,\,\,\,\,\,\,\,\,\, \label{KGSS}
\end{eqnarray}
Here, $\hbar$, $c$, and $m_0$  represent the Planck constant, the speed of light, and the rest mass of the confined particle, respectively. Then, we substitute the GSWSP energy given in  Eq.~(\ref{GSWSP}) into the KGE and we obtain
  \begin{eqnarray}
  \Bigg[\frac{d^2}{dx^2}+\alpha^2\bigg[-\epsilon^2+\frac{\beta^2}{1+e^{-\alpha(x+L)}}+\frac{\gamma^2}
  {\big(1+e^{-\alpha(x+L)}\big)^2}\bigg] \Bigg]\phi_L(x) &=& 0, \label{KG2x<0}
\end{eqnarray}
for  the $x<0$ region with the given abbreviations
\begin{eqnarray}
-{\epsilon}^2 &\equiv& \frac{E^2-\big(\varepsilon^{+}- m_0c^2\big)^2}{\alpha^2\hbar^2c^2},\\
{\beta}^2  &\equiv& \frac{2\Big(E- \big(\varepsilon^{+}- m_0c^2\big)\Big)(V_0-W)}{\alpha^2\hbar^2c^2},  \\
{\gamma}^2    &\equiv& \frac{2\Big(E- \big(\varepsilon^{+}-m_0c^2\big)\Big)W}{\alpha^2\hbar^2c^2}.
\end{eqnarray}
Even though, here we closely follow our previous paper \cite{RefLutfuogluEPJP2018}, there is a remarkable difference. We keep the differentiation parameter in the equations in order to investigate its role on the spectrum. By defining a new
transformation
\begin{eqnarray}
  z &\equiv& \Big[1+e^{-\alpha(x+L)}\Big]^{-1},
\end{eqnarray}
we derive a dimensionless equation from Eq. (\ref{KG2x<0})
\begin{eqnarray}
&&\Bigg[\frac{d^2}{dz^2}+\bigg(\frac{1}{z}+\frac{1}{z-1}\bigg)\frac{d}{dz}+\bigg(\frac{\beta^2-2\epsilon^2}{z}
-\frac{\epsilon^2}{z^2}-\frac{\beta^2-2\epsilon^2}{(z-1)}+\frac{\beta^2+\gamma^2-\epsilon^2}
{(z-1)^2}\bigg)\Bigg]  \phi_{L}(z) = 0. \,\,\,\,\,\,\, \label{KG3x<0}
\end{eqnarray}
As an ansatz, we propose the general solution in the form of
\begin{eqnarray}
  \phi_{L}(z) &\equiv& z^{\mu} (z-1)^{\nu}f(z) \label{gensol}
\end{eqnarray}
where, $\mu^2-\epsilon^2=0$  and $\nu^2+\beta^2+\gamma^2-\epsilon^2 = 0$. Furthermore, we define positive wave numbers $k$ and $\kappa$
\begin{eqnarray}
k&\equiv&\frac{1}{\hbar c}\sqrt{-\big(E- \varepsilon^{+}+ m_0c^2\big)\big(E+ \varepsilon^{+}- m_0c^2 \big)}\,,\label{k1} \\
\kappa &\equiv& \frac{1}{\hbar c} \sqrt{\big(E- \varepsilon^{+}+ m_0c^2\big)\big(E+ \varepsilon^{+}- m_0c^2+2V_0 \big)}\,. \label{k2}
\end{eqnarray}
that satisfy
\begin{eqnarray}
  \mu  &=& \frac{k}{\alpha}, \\
  \nu  &=& \frac{i\kappa}{\alpha}.
\end{eqnarray}
The resulting equation is the  Hypergeometric differential equation
\begin{eqnarray}
   &&z(1-z)f{''}+ \Big[(1+2\mu)-(1+2\mu+2\nu+1)z\Big]f{'}-\Big[(\mu+ \nu)^2+(\mu+\nu)+\gamma^2\Big]f= 0, \label{KG_eq5}
\end{eqnarray}
that possesses solutions in terms of ${}_2F_1$, which is a hypergeometric function. Finally, we derive the general solution in $x<0$ region in the following form
\begin{eqnarray}
  \phi_L(z) &=& D_1 z^{\mu} (z-1)^{\nu} \,\,\, {}_2F_1[\mu+\theta+\nu, 1+\mu-\theta+\nu, 1+2\mu; z]\nonumber \\
  &+&D_2 z^{-\mu} (z-1)^{\nu} \,\,\, {}_2F_1[-\mu+\theta+\nu, 1-\mu-\theta+\nu, 1-2\mu; z]. \,\,\,\,\,\,\,\,
\end{eqnarray}
where $\theta$ is defined to be
\begin{eqnarray}
\theta \equiv \frac{1}{2}\mp \sqrt[]{\frac{1}{4}-\gamma^2}\,.
\end{eqnarray}
Note that, from now on, we prefer to use the "minus definition" in our calculation. As a consequence of the symmetry, $x\longrightarrow-x$,  of the GSWSP energy, we obtain a symmetric solution in region $x>0$.
\begin{eqnarray}
  \phi_R(y) &=& D_3 y^{\mu} (y-1)^{\nu} \,\,\, {}_2F_1[\mu+\theta+\nu,1+\mu-\theta+\nu,1+2\mu;y]\nonumber \\
  &+&D_4 y^{-\mu} (y-1)^{\nu} \,\,\, {}_2F_1[-\mu+\theta+\nu,1-\mu-\theta+\nu,1-2\mu;y].
\end{eqnarray}
Note that, in the positive region we use the transformation
\begin{eqnarray}
  y \equiv \Big[1+e^{\alpha(x-L)}\Big]^{-1}.
\end{eqnarray}
As a final remark, we denote the normalization constants with $D_1$, $\cdots$, $D_4$.

\subsection{Bound State Conditions}\label{Bound State Conditions}
In a bound state problem, the boundary condition predicts that the particle's wave function has to vanish exponentially outside the well. This condition is satisfied if and only if, $\mu$ is a real number, whereas, $\nu$ is an imaginary number. We assign these constraints on the  wave numbers defined in  Eq.~(\ref{k1}) and  Eq.~(\ref{k2}). Then, we obtain the conditions
\begin{eqnarray}
   -\big(E- \varepsilon^{+}+ m_0c^2\big)\big(E+ \varepsilon^{+}- m_0c^2 \big) &>&0, \label{SSbirincikosul}\\
  \big(E- \varepsilon^{+}+ m_0c^2\big)\big(E+ \varepsilon^{+}- m_0c^2+2V_0 \big)&>&0, \label{SSikincikosul}
\end{eqnarray}
besides to the condition $V_{cr}>V_0>0$. Here, $V_{cr}$ value is an upper limit of the potential depth parameter due to the Klein paradox \cite{RefCalogeracosDombey1999}.

\begin{figure}[!htb]
\centering
\includegraphics[totalheight=0.45\textheight,clip=true]{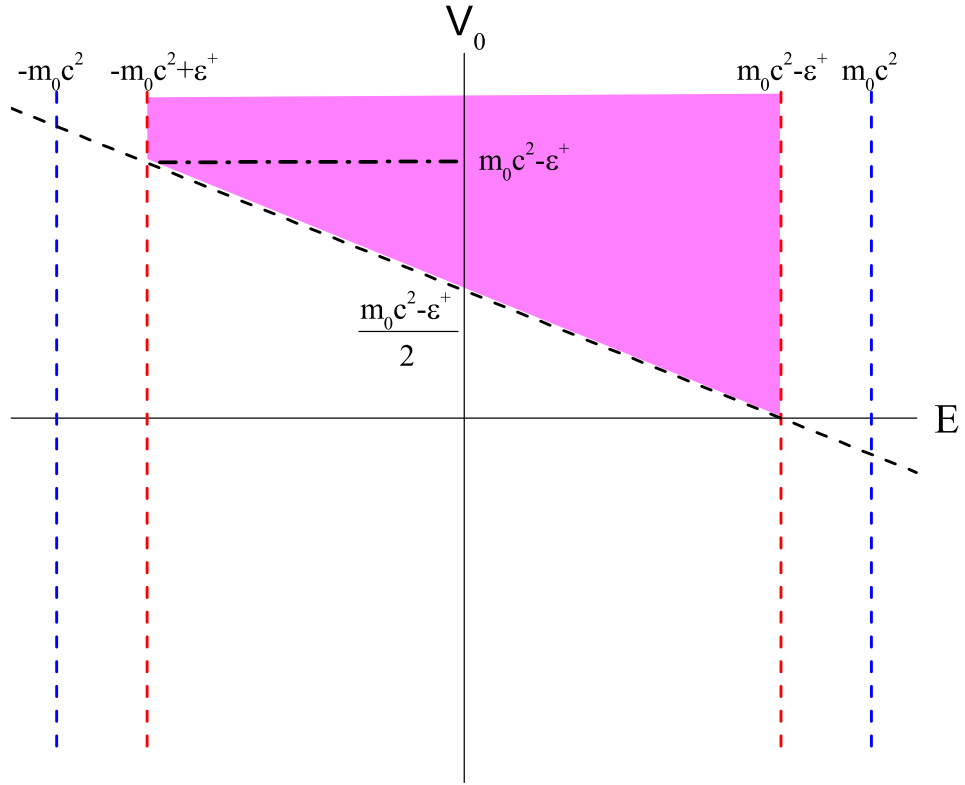}
   \caption{Possible energy eigenvalue region for a confined particle in  SS limit. Potential well depth parameter plays a crucial role in order to have positive and/or negative eigenvalues. } \label{fig6:SSpotandenergyrelations}
\end{figure}

We investigate these conditions comprehensively in order to comprehend the crucial role of the differentiation parameter. In  Fig.~\ref{fig6:SSpotandenergyrelations}, we display them by representing the parameters $V_0$ and $E$ on the axes. The shaded area  indicates the intersection of the required constraints.

We observe that the first constraint, given in  Eq.~(\ref{SSbirincikosul}), circumscribes the energy spectrum in a limited interval.

The second constraint, given in  Eq.~(\ref{SSikincikosul}) determines the possible minimum value limit of the energy spectrum. For instance, the possibility of the occurrence of an energy spectrum  that has values only greater than zero depends on the potential depth parameter to be less than $\frac{m_0c^2-\varepsilon^+}{2}$.
If the potential depth parameter has a value in the interval of $(0, m_0c^2-\varepsilon^+ )$, then negative values can be obtained in the energy spectrum with constriction. The minimum value limit of the spectrum depends on the value of the depth parameter. Off the greater values of the depth parameter, until a critical value, the energy spectrum occurs with the all the possible values that are allowed from the first constraint given in Eq.~(\ref{SSbirincikosul}). Therefore, we conclude that one of the role of the differentiation parameter is to modify the mass energy of the confined particle and consequently to narrow the spectrum interval.

Before we apply the continuity conditions, we examine the behavior of wave functions at positive and negative infinities. We find
\begin{eqnarray}
 \phi_L(x\rightarrow -\infty) &\approx& \Big[D_1 e^{x k } + D_2^\mp e^{-xk } \Big]e^{-\frac{\pi \kappa}{\alpha}} , \\
 \phi_R(x\rightarrow \infty) & \approx & \Big[D_3 e^{-xk} + D_4  e^{xk } \Big]e^{-\frac{\pi \kappa}{\alpha}}.
\end{eqnarray}
We assume that the name numbers $k$ and $\kappa$ are positive numbers. This assumption causes two of the normalization constants, $D_2$ and $D_4$, to be equal to zero.
\subsection{The Continuity Conditions}\label{CC}
If the potential energy has finite values, the continuity condition of the derivative of the wave function is accompanied by the condition of the continuity of the wave function. Therefore, in this study at the critical point $x=0$, the wave function and its derivative should be equal to each other  in the positive and negative regions.

After simple and straightforward calculations we obtain two equations that correspond to the continuity of the wave function
\begin{eqnarray}
(D_1-D_3)t_0^{\mu}(t_0-1)^{\nu}M_1&=&0, \label{kendisi}
\end{eqnarray}
and to the continuity of the derivative of the wave function
\begin{eqnarray}
(D_1+D_3)t_0^{\mu}(t_0-1)^{\nu}\Bigg[\Bigg(\frac{\mu}{t_0}+\frac{\nu}{t_0-1}\Bigg)M_1+
\frac{(\mu+\theta+\nu)(1+\mu-\theta+\nu)}{1+2\mu} M_3\Bigg] = 0 . \label{turevi}
\end{eqnarray}
Here, we define
\begin{eqnarray}
  t_0 \equiv (1+e^{-\alpha L})^{-1},
\end{eqnarray}
and use new abbreviations
\begin{eqnarray}
  M_1 &=& S_1 N_1+ (1-t_0)^{-2\nu}S_2 N_2, \\
  M_3 &=& S_3 N_3+ (1-t_0)^{-1-2\nu}S_4 N_4,
\end{eqnarray}
where
\begin{eqnarray}
  S_1 &\equiv& \frac{\Gamma(1+2\mu)\Gamma(-2\nu)}{\Gamma(1+\mu-\theta-\nu)\Gamma(\mu+\theta-\nu)}, \\
  S_2 &\equiv& \frac{\Gamma(1+2\mu)\Gamma(2\nu)}{\Gamma(1+\mu-\theta+\nu) \Gamma(\mu+\theta+\nu)}, \\
  S_3 &\equiv& \frac{\Gamma(2+2\mu)\Gamma(-1-2\nu)}{\Gamma(1+\mu-\theta-\nu)\Gamma(\mu+\theta-\nu)}, \\
  S_4 &\equiv& \frac{\Gamma(2+2\mu)\Gamma(1+2\nu)}{\Gamma(2+\mu-\theta+\nu)\Gamma(1+\mu+\theta+\nu)},
\end{eqnarray}
and
\begin{eqnarray}
  N_1 &\equiv&  \,\,\, {}_2F_1[\mu+\theta+\nu,1+\mu-\theta+\nu,1+2\nu;1-t_0],  \\
  N_2 &\equiv&  \,\,\, {}_2F_1[1+\mu-\theta-\nu,\mu+\theta-\nu,1-2\nu;1-t_0],  \\
  N_3 &\equiv&  \,\,\, {}_2F_1[1+\mu+\theta+\nu,2+\mu-\theta+\nu,2+2\nu;1-t_0],  \\
  N_4 &\equiv&  \,\,\, {}_2F_1[1+\mu-\theta-\nu,\mu+\theta-\nu,-2\nu;1-t_0].
\end{eqnarray}

\subsection{Quantization and the Wave Function Solutions}\label{Quan}

We use the continuity conditions, derived in  Eq.~(\ref{kendisi}) and Eq.~(\ref{turevi}), to obtain the energy spectrum. In one dimension, due to the current symmetry, we divide the energy spectrum into two subsets in means of odd and even node numbers.

We obtain the even subset of the energy spectrum, $E_n^e$, by setting $D_1=D_3$.  Eq.~(\ref{kendisi}) becomes identically equal to zero. The other equation,  Eq.~(\ref{turevi}), yields to

\begin{eqnarray}
  \frac{(S_1 N_1)+(1-t_0)^{-2\nu}(S_2 N_2)}{(S_3 N_3)+(1-t_0)^{-1-2\nu}(S_4 N_4)}&=&  -\frac{(\mu+\theta+\nu)(1+\mu-\theta+\nu)t_0(t_0-1)}
{(1+2\mu)\big((\mu+\nu)t_0-\mu\big)}. \label{ciftcozum}
\end{eqnarray}
Furthermore, we obtain the even wave function in two parts that corresponds to the negative and positive regions, as follows:
\begin{eqnarray}
  \phi_L(z) &=& D_1 z^{\mu} (z-1)^{\nu} \,\,\, {}_2F_1[\mu+\theta+\nu,1+\mu-\theta+\nu,1+2\mu;z],\\
  \phi_R(y) &=& D_1 y^{\mu} (y-1)^{\nu} \,\,\, {}_2F_1[\mu+\theta+\nu,1+\mu-\theta+\nu,1+2\mu;y].
\end{eqnarray}

On the other hand, we establish the odd subset of the energy spectrum, $E_o^e$, by setting $D_1=-D_3$, hence Eq.~(\ref{turevi}) becomes to be equal to zero. Moreover,  Eq.~(\ref{kendisi}) gives
\begin{eqnarray}
  \frac{S_1 N_1}{S_2 N_2} &=& -(1-t_0)^{-2\nu}. \label{tekcozum}
\end{eqnarray}
Alike the even case, we obtain the odd wave function in two parts that corresponds to negative and positive regions, respectively.
\begin{eqnarray}
  \phi_L(z) &=& D_1 z^{\mu} (z-1)^{\nu} \,\,\, {}_2F_1[\mu+\theta+\nu,1+\mu-\theta+\nu,1+2\mu;z],\\
  \phi_R(y) &=& -D_1 y^{\mu} (y-1)^{\nu} \,\,\, {}_2F_1[\mu+\theta+\nu,1+\mu-\theta+\nu,1+2\mu;y].
\end{eqnarray}

\section{Results and Discussions} \label{sec:uygulama}
In this section we employ the NR numerical methods. We assume that the confined particle in the GSWSP energy well is a neutral Kaon. Note that its rest mass energy is $m_0c^2=497.648$ \,\,$MeV$. We suppose that the energy well is constructed with  $V_0=\frac{m_0c^2}{2}$, $\alpha=1\,\, fm^{-1}$ and $L=6\,\, fm$ parameters in addition to the  $W=2 m_0c^2$  and  $W=-2 m_0c^2$  parameters in the repulsive and attractive surface effects cases, respectively.

\subsection{The role on the bound state solutions with repulsive surface effects}

To calculate the energy spectra at various values of differentiation parameter, we use the equations obtained in Eq.~(\ref{ciftcozum}) and Eq.~(\ref{tekcozum}). We tabulate the calculated energy spectra in Table~\ref{tab1} with their corresponding node numbers, $n$. For the various differentiation parameters, we calculate the ratio of the calculated energy spectra to the bound particles' rest mass energy, and then, we plot these ratios with the number of nodes in Fig.~\ref{fig8:KGboundEnergies}.

In Fig.~\ref{fig9:potenerji-2}, we examine the role of the differentiation parameter on the energy spectrum. The first bound state conditions, given in Eq.~(\ref{SSbirincikosul}), states that the Klein-Gordon interval should be in between the points $A$ and $I$. The second condition, given with Eq.~(\ref{SSikincikosul}), constricts the interval to be among the points $F$ and $I$. This analysis is in a complete agreement with the calculated energy spectrum, that possesses only positive values, as given in the second column of Table~\ref{tab1}. Note that the allowed interval length is equal to $m_0c^2$.

In the case of bound state problems, the depth parameter of the potential energy well in which the bound particle is located is initially determined and is constant throughout the problem. The increase of the differentiation parameter plays the role of the shift of the energy interval, due to the conditions that are given in Eq.~(\ref{SSikincikosul}). We demonstrate this effect in  Fig.~\ref{fig9:potenerji-2}. For instance, when $\varepsilon^+=50$ $MeV$,  $F$ and $I$ points shift to the points $D$ and $H$, respectively. The consequence of such a shift the energy spectrum could have negative eigenvalues. In this particular problem, we observe that with the increase of the exchange parameters, the ground state energy eigenvalue approaches to zero and possesses a negative value when $\varepsilon^+=40$ $MeV$, as predicted. We would like to remark that the shifted energy interval's length does not change via the increase of the differentiation parameter.

Next, we assign an extreme value to the differentiation parameter. When  the differentiation parameter is equal to $\frac{m_0c^2}{2}$, the shifted Klein-Gordon interval  becomes symmetric  in between $C$ and $G$.  Although the interval length remains constant, we obtain a decrease in the number of the calculated eigenvalues in the energy spectrum. Therefore, we conclude that the differentiation parameter has an effect on the eigenvalue density of the energy spectrum.

Finally, in Fig~\ref{fig10:degisim} we investigate the role of the differentiation parameter on the energy step size in their energy spectra. Here, we name the difference between any two consecutive eigenvalues as an energy step size. In each value of the differentiation parameter, the energy step size landscape goes uphill until three increments of node number, then a decrease follows with further increment in node number. When the differentiation parameter exceeds to $\frac{m_0c^2}{2}$, the energy step size only decreases with node number increment, but this is not illustrated in Fig~\ref{fig10:degisim}. The increase in the differentiation parameter gives rise to the increase in the energy step size in the constant interval, therefore, the number of available energy eigenvalues tends to decrease after a critical value. A decrease in the number of nodes until $\varepsilon^+=50$ $MeV$ is not observed, whereas, for an asymptotic value of the differentiation parameter, chosen as $\frac{m_0c^2}{2}$, the number of nodes decreases from nine to seven. As a final remark, we observe that at the very end of the increment in node number scale, versus the differentiation parameter, the landscape almost flattens with very small step sizes. Contrarily, at the beginning of the increment in node number scale, versus the differentiation parameter, the landscape increases at a relatively high rate as we demonstrate in Fig~\ref{fig10:degisim}.

\subsection{The role on the bound state solutions with attractive surface effects}

In this subsection, we employ the NR method to solve Eq.~(\ref{ciftcozum}) and Eq.~(\ref{tekcozum}) for $\varepsilon^+=0$ and $\varepsilon^+=50\,\,MeV$ under the presence of attractive surface effects. In Table~\ref{tab2}, we tabulate the calculated energy spectra versus the node numbers. When we choose $\varepsilon^+=0$, we obtain sixteen eigenvalues lowest one's node number is six. However, when we take  $\varepsilon^+=50$ $MeV$, we find the wave function corresponding to the lowest energy eigenvalue in the spectrum has four node number. In addition, we get seventeen eigenvalues in the energy spectrum.

In Fig.~\ref{fig11:degisim}, we plot the node numbers versus the rate of the energy spectra to the rest mass energy of neutral Kaon. We observe that the rate becomes negative with the increase in the parameter $\varepsilon^+$, similar to the behavior  in the repulsive case. On the other hand, the number of energy eigenvalues in the spectrum increase, in contrast with the repulsive case.

\section{Conclusion}\label{sec:sonuc}

In this manuscript, we investigate the role of the differentiation parameter on the solution of the KGE in the SS limit by arising a novel analogy of the variational method. We examine the bound state conditions under the GSWSP energy with the presence of the differentiation parameter. Then, we discuss the continuity condition and quantization scheme. We observe that one  of the role of the differentiation parameter is to modify the confined particle's mass-energy, hence to narrow the spectrum interval length. In order to obtain numerical results, we use a neutral Kaon confinement in a GSWSP energy well. In the presence of differentiation parameters, we obtain various energy spectra for the repulsive and attractive surface effect cases. As consequences of discussions that have been done, we conclude that  the differentiation parameter plays the role of to be a measure of the density of the eigenvalues. Furthermore, we show that higher values of differentiation parameter correspond to a greater step size in the spectrum compared to  that of its lower values, in both repulsive and attractive surface effects.

\section*{Acknowledgments}
This work was partially supported by the Turkish Science and Research Council \,\, (T\"{U}B\.{I}TAK) and Akdeniz University. The author thanks for the support given by the Internal Project of Excellent Research of the Faculty of Science of University Hradec Kr\'{a}lov\'{e}, "Studying of properties of confined quantum particle using Woods-Saxon potential".  The author is indebted to Prof. M. Horta\c{c}su and Prof. J. K\v{r}\'{i}\v{z} for the proof reading. The author thanks to the reviewers for their kind recommendations that leads several improvements in the article.


\section{References}

\newpage
\begin {table}[!ht]
\caption{\label{tab1}
Energy spectra for different values of the differentiation parameter for the repulsive surface interaction case. Note that all calculated eigenvalues and $\varepsilon^+$ have units in $MeV$.}
\begin{tabular}{|l|c|c|c|c|c|c|c|}
  \hline
  n / $\varepsilon^+$ & 0& 10 & 20  & 30 & 40 & 50 & $\frac{m_0c^2}{2}$     \\\hline
  0 & 33.962  & 24.497  & 15.065  & 5.667   & -3.691  & -13.005 & -166.575  \\
  1 & 86.193  & 77.627  & 69.133  & 60.717  & 52.384  & 44.141  &  -80.716  \\
  2 & 139.950 & 132.155 & 124.452 & 116.849 & 109.350 & 101.962 &   -6.487  \\
  3 & 194.935 & 187.760 & 180.686 & 173.718 & 166.863 & 160.125 &   61.947  \\
  4 & 249.319 & 242.635 & 236.055 & 229.582 & 223.222 & 216.978 &  125.727  \\
  5 & 303.138 & 296.845 & 290.653 & 284.567 & 278.590 & 272.726 &  186.228  \\
  6 & 355.809 & 349.826 & 343.942 & 338.159 & 332.481 & 326.910 &  243.686  \\
  7 & 407.437 & 401.701 & 396.059 & 390.513 & 385.066 & 379.721 &  \\
  8 & 457.746 & 452.203 & 446.749 & 441.386 & 436.116 & 430.941 &  \\
  \hline
\end{tabular}
\end{table}

\begin {table}[!ht]
\caption{\label{tab2}
Energy spectra for different values of the differentiation parameter for attractive surface forces. Note that all calculated energies and $\varepsilon^+$ have units in $MeV$.}
\begin{tabular}{|l|c|c|l|c|c|}
  \hline
  n / $\varepsilon^+$   &    0    & 50      & n / $\varepsilon^+$   & 0       & 50        \\\hline
  4                     &         & -40.169 & 13                    & 274.238 & 256.673   \\
  5                     &         & -14.710 & 14                    & 305.153 & 283.845   \\
  6                     & 47.403  & 18.052  & 15                    & 335.082 & 313.494   \\
  7                     & 78.700  & 51.906  & 16                    & 371.316 & 341.792   \\
  8                     & 111.445 & 86.490  & 17                    & 391.456 & 368.410   \\
  9                     & 144.380 & 120.811 & 18                    & 417.433 & 393.020   \\
  10                    & 177.494 & 154.910 & 19                    & 441.465 & 414.986   \\
  11                    & 210.232 & 188.343 & 20                    & 463.049 & 433.286   \\
  12                    & 242.588 & 221.124 & 21                    & 481.232 &           \\
  \hline
\end{tabular}
\end{table}

\newpage

\newpage
\begin{figure}[!htb]
\centering
\includegraphics[totalheight=0.45\textheight,clip=true]{fig2.eps}
   \caption{Node numbers versus the rate of the bound state's energy spectra to the rest mass energy of Kaon particle for  different values of $\varepsilon^+$ in the existence of the repulsive surface effects. Note that $0.5 m_oc^2=248.824$ $MeV$.} \label{fig8:KGboundEnergies}
\end{figure}

\newpage
\begin{figure}[!htb]
\centering
\includegraphics[totalheight=0.45\textheight,clip=true]{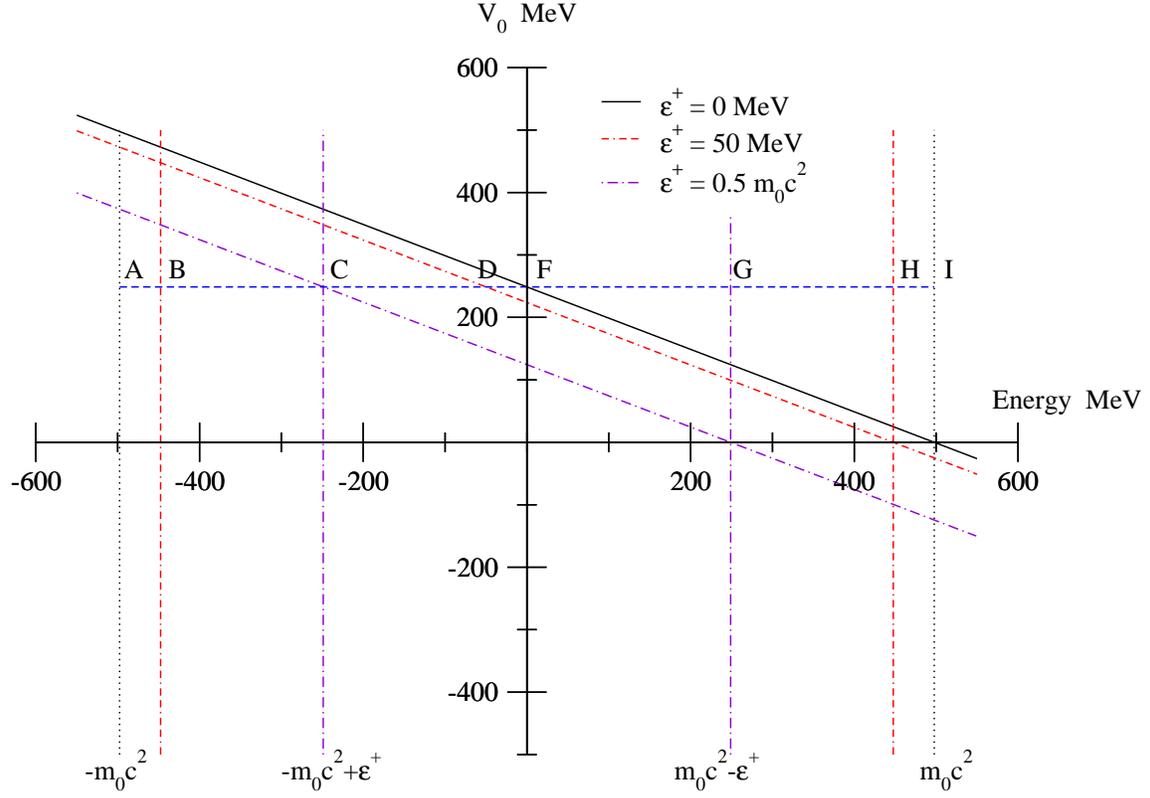}
   \caption{The change of bound state energy spectrum interval for a fixed value of potential depth parameter versus $\varepsilon^+=0$.  When $\varepsilon^+=0$, the energy spectrum is constituted with the values of energy between $F$ to $I$ points. While the $\varepsilon^+=0$ increase the Klein-Gordon energy interval shifts to be $D$ to $H$. Note that the spectrum interval gains a symmetry  when $\varepsilon^+$ takes the half value of the neutral Kaon mass energy.}  \label{fig9:potenerji-2}
\end{figure}

\newpage
\begin{figure}[!htb]
\centering
\includegraphics[totalheight=0.35\textheight,clip=true]{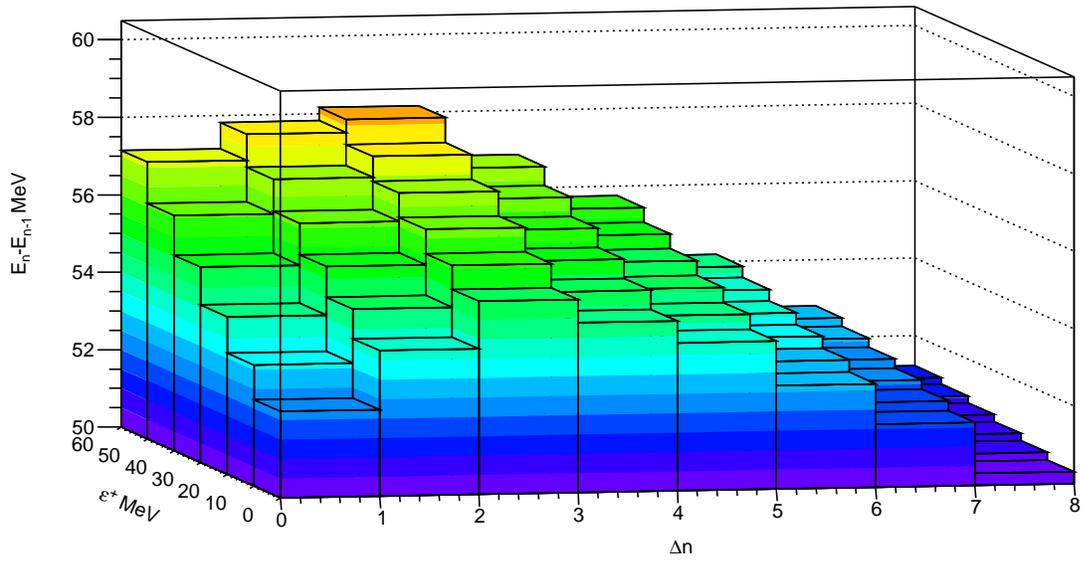}
   \caption{Three dimensional diagram among the increment of node numbers, the differentiation parameter and energy step size.} \label{fig10:degisim}
\end{figure}

\newpage
\begin{figure}[!htb]
\centering
\includegraphics[totalheight=0.40\textheight,clip=true]{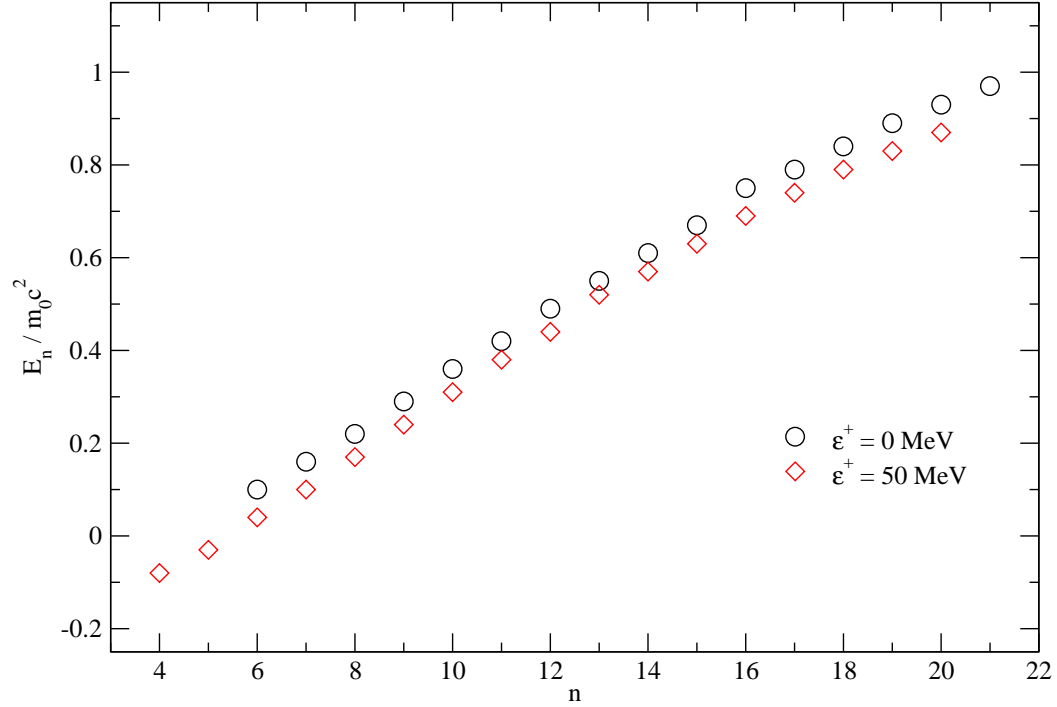}
   \caption{Node numbers versus the rate of the bound state's energy spectra to the rest mass energy of Kaon particle for  different values of the differentiation parameter in the existence of the attractive surface effects. Note that $0.5 m_oc^2=248.824$ $MeV$.} \label{fig11:degisim}
\end{figure}

\end{document}